\pdfoutput=1
\documentclass[nofootinbib,twocolumn,showpacs,amssymb,floatfix,deluxe,aps]{revtex4}

\usepackage [latin1]{inputenc}
\usepackage{graphicx}
\usepackage{dcolumn}
\usepackage{bm}
\usepackage{epsfig}
\usepackage[english]{babel}
\usepackage{amsmath}
\usepackage{amssymb}

\usepackage{verbatim}
\usepackage{alltt}

\def\({\left(}
\def\){\right)}

\begin{document}


\preprint{}
\title{
The Cosmic-Ray Spectra: \\
News on their Knees} 

\author{A. De R\'ujula${}^{a,b}$}
\affiliation{  \vspace{3mm}
${}^a$Instituto de F\'isica Te\'orica (UAM/CSIC), Univ. Aut\'onoma de Madrid, Spain;\\
${}^c$Theory Division, CERN, CH 1211 Geneva 23, Switzerland
}

\date{\today}

\begin{abstract}

In a comprehensive model of Cosmic Rays (CRs) proposed a decade ago, the energies of the spectral 
``knees" of the various CR species were predicted to be proportional to mass, rather than charge.
The model also predicts the knees to occur at an energy of two to four million times the particle's
rest mass. Recent data allow one to verify this prediction, particularly for Fe and lighter-nuclei
 CRs. But the most stringent 
test involves the putative knee in the CR electron spectrum, since the mass ratio of electrons to protons
(and nuclei) is so very different from their charge ratio(s). Very recent results on the spectra of positrons
and electrons at the highest measured energies corroborate the existence of an electron knee, with the
expected shape and at the predicted energy. 
\end{abstract}


\maketitle

\section{Introduction}

Cosmic Rays (CRs) occupy a very peculiar niche in physics. Though they were discovered more than 
a century ago, there is no acceptable and accepted theory that describes them. This is in spite of the fact
that there is no reason to believe that the understanding of CRs 
would require any revolutionary ingredients. And this applies
 not only to protons and other nuclei, but also to CR electrons and positrons, for which
claims of ``physics beyond the standard model" abound.

A significant contribution to the lack of a generally trusted CR theory is that the 
number of correct predictions in the field is impressively small. One exception is
the prediction by Giuseppe Cocconi and Philip Morrison \cite{Cocconi,Morrison}
that the magnetic field of the galaxy would not be able to confine
CRs of energy exceeding ${\rm Z}\,(3\times 10^9)$ GeV, with Z the CR's charge.
Some spectral feature is then expected at such an energy. This, for $\rm Z=1$, turns out
to be seen as the ``ankle" in the all-particle spectrum. Auger has recently observed a
dipolar asymmetry in the incoming directions of CRs of energy above $8\times 10^9$ GeV,
pointing in the sky in a direction very different to that of the galactic center \cite{Auger}.
Cocconi and Morrison were right.

Another beautifully simple prediction is the ``GZK cutoff" at energies exceeding 
${\rm A}\,(5\times 10^{10})$ GeV, with A the CR's mass number \cite{GZK}. Ultra-High
Energy Cosmic Rays with energies amply exceeding the proton's GZK limit 
are observed and their composition is unknown. 
The original GZK cutoff reflects the pion-production threshold on the cosmic microwave and infrared backgrounds. The understanding of the highest-energy CR flux is complicated by the fact that photo-dissociation of primary CR nuclei is relevant at similar energies.

A decade ago a Cannon-Ball (CB) model of CRs \cite{DP} 
was elaborated \cite{DD2008}. The model is very economic,
in the sense of having only one free parameter to be fit to the data. Its remaining degrees of
freedom concern ``priors", information to be gathered from observations independent from
the model. With choices of the priors in their allowed range, the model was
shown to accurately describe all properties of (primary, non-solar) CRs. New data allow one to
discuss a prediction of the CB model, concerning the knees in the
spectra of the fluxes of individual nuclei and electrons. The prediction turns out
to be right.

The CB model of Gamma-Ray Bursts and X-Ray Flashes \cite{SD,GRB1,DD,AGoptical,AGradio,DDDXRF} 
(collectively, GRBs), CRs,
the gamma background radiation \cite{DDGBR}, cooling flows \cite{CDD} and neutron-star 
mergers \cite{NS}
cannot be accused of being unjustifiably popular.
For that reason, I summarize in an appendix the information required to understand
the basis for the predictions of the CB model concerning the CR knees.

Suffice it to say at this point that core-collapse supernovae (SNe) produce highly relativistic 
jets of ``cannonballs" of ordinary matter. These CBs, colliding with the constituents
of the interstellar medium (ISM), promote them to CR energies. GRBs
and their afterglows are also generated by CBs launched by SNe, when accurately
pointing to the
observer from a distant galaxy. The successful 
CB-model description of GRBs allows one to extract the information
required to predict the properties of CRs.

\section{The Knees}

\begin{figure}[]
\vspace{-4cm}
\hspace{-1.5cm}
\epsfig{file=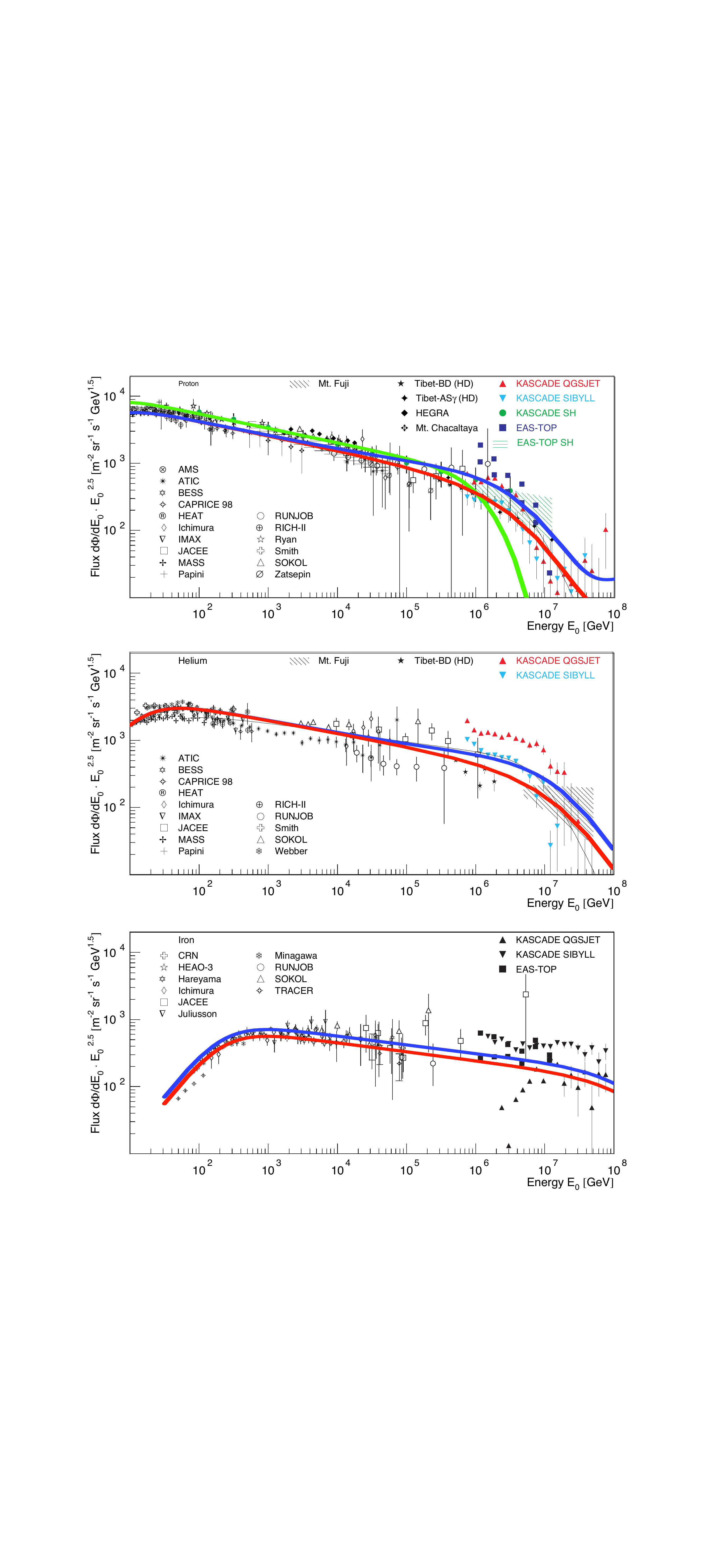,width=10cm}
\vspace{-5,5cm}
\caption{Early data \cite{Hor} on the p, He and Fe spectra (times $E^{2.5}$) 
 and their early CB-model description \cite{DD2008}.
The red and blue curves
indicate the sensitivity to contributions subdominant in the plotted energy range,
with their priors chosen within their ranges.
The green curve illustrates the choice of a very narrow distribution of CB's initial
Lorentz factors. The compilation of data was at the time kindly provided to us by K.H. Kampert.}
\label{fig:OldKnees}
\end{figure}

About a decade ago the data on the separate spectra of protons and He and Fe nuclei
were the ones depicted in Fig.~(\ref{fig:OldKnees}). Also shown in this figure are various
CB-model descriptions of the data, corresponding to choices of the priors within their
respective ranges \cite{DD2008}. These data showed a significant knee for protons, and an indication of
a He knee. The measurements did not extend high enough in energy to reflect
a potential Fe knee.


In the range of energies shown in Fig.~(\ref{fig:OldKnees}) the dominant contribution to the
CB-model spectra is the scattering by a moving CB of the constituents of the ISM, 
previously ionized by the GRB's $\gamma$ rays.  The red and blue curves
indicate the sensitivity to subdominant contributions: the extragalactic CR flux and the CRs
having been accelerated in a CB's inner magnetic field. These small 
contributions are neglected in what follows; they have an inconsequential 
 impact on the discussion of the spectral knees, on which we are interested here. 

The curves in Fig.~(\ref{fig:OldKnees}) reflect the fact that the CB model correctly describes the
CR data at all energies including, though not shown in the figure, the data at the highest
and smallest measured energies. In the low-energy domain the
CB-model description of the spectra is not as simple as for relativistic energies  \cite{DD2008}.
For the sake of expedience in discussing the knees, I shall only
use here the CB-model's results for very relativistic CRs.

\begin{figure}[]
\centering
\vspace{.5cm}
\epsfig{file=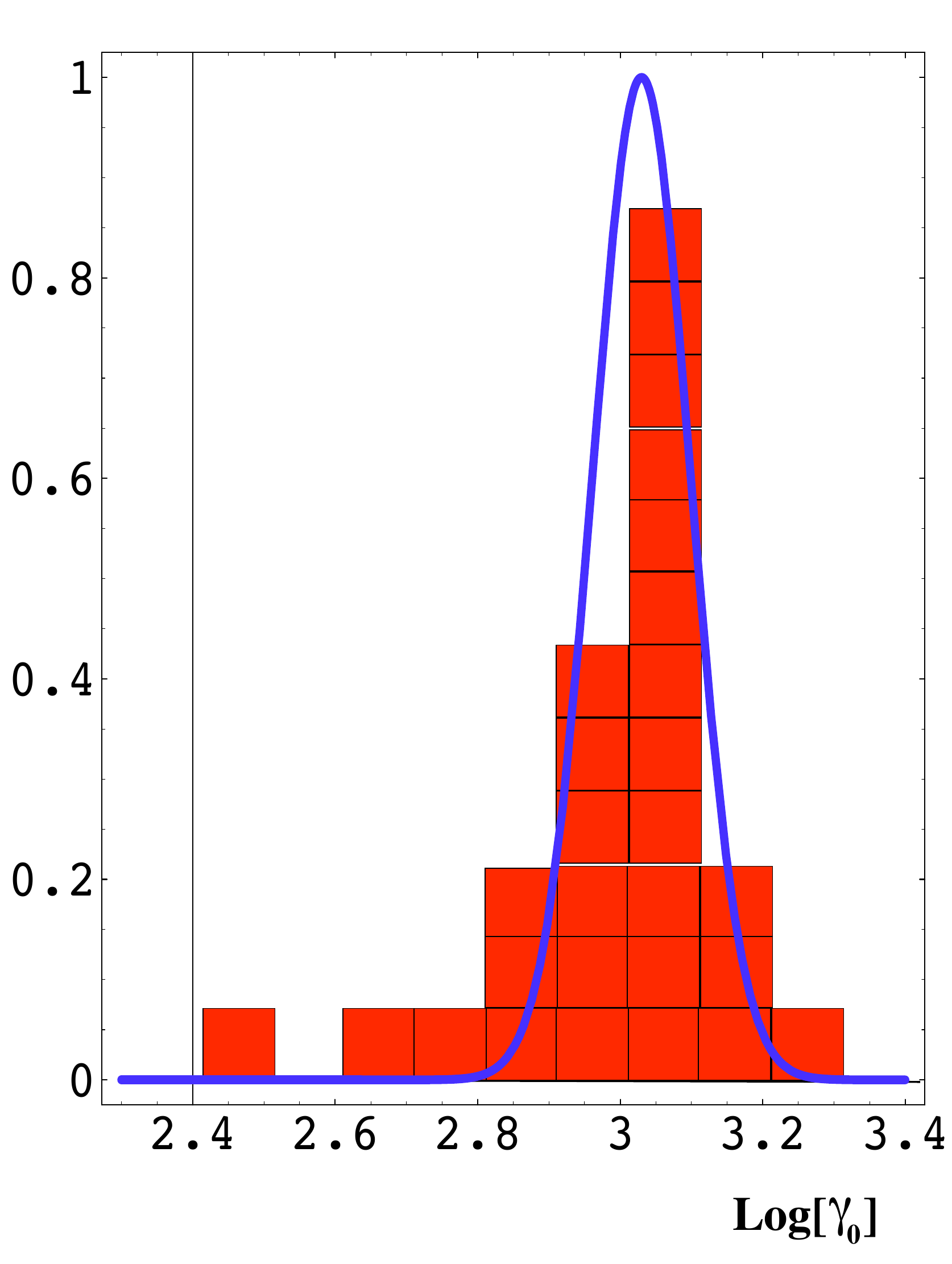,width=6cm}
\caption{The distribution $D(\gamma_0)$ of initial Lorentz factors of CBs, extracted from data
on GRBs and their afterglows \cite{DD2008}. The blue curve is a log-normal description of the data.}
\label{fig:gammaDist}
\end{figure}

Let $\gamma_0$ be the initial Lorentz factor (LF) with which a given CB is ejected in a supernova
event. A distribution of $\gamma_0$ values, $D(\gamma_0)$ --extracted from the analysis of GRBs 
and their afterglows-- is shown in Fig.~(\ref{fig:gammaDist}). It peaks at $\bar \gamma_0$ slightly
larger than $10^3$, above which it falls abruptly. The data at $\gamma_0<\bar \gamma_0$
may be under-represented due to observational selection effects (smaller GRB fluxes), 
but only the high-$\gamma_0$ part of $D(\gamma_0)$ is relevant to the location and shape
of the CRs knees.

Soon after its launching, a CB expands to a point where its density is low enough for
it to become transparent --in the sense of
its individual constituents and the ones of the ISM it encounters not to scatter significantly.
But a CB has an inner turbulent magnetic field, as explained in the Appendix.
This field captures and scatters the charged ISM constituents. As the
CB expands and slows down in its trajectory, this results
in the ISM particles being re-emitted by the CB
as CRs with a distribution of their LFs, $\gamma_{_{\rm CR}}$:
\begin{equation}
{dF\over d\gamma_{_{\rm CR}}}\propto n_{_{\rm A}}\,\gamma_{_{\rm CR}}^{-\beta_s}\,
\Theta[2\,\gamma_0^2-\gamma_{_{\rm CR}}],\;\;\; \beta_s={13\over 6}\approx 2.17,
\label{eq:source}
\end{equation}
\noindent
with $n_{_{\rm A}}$ the abundance of the ISM nuclei of nucleon number A that the CB collides
with \cite{DD2008}.

The upper limit $\gamma_{_{\rm CR}}\leq 2\,\gamma_0^2$ is easy to 
understand\footnote{ The derivation of the spectral index $\beta_s$ is elaborate. It was
once assessed by a cunning referee as ``almost Baron Munchhausen", presumably
meaning a lengthy and ingenious list of fabrications. If I may,
I would substitute ``fabrications" for ``arguments".}.
Consider a freshly ejected CB in its rest frame. The incoming ISM particles reach it with
a LF $\gamma_0$. The ones elastically back-scattered have in the CB's rest frame
a LF $\gamma_0$, the mass of the CB being so much larger than
the energy of the ISM projectile.
Lorentz-boost these scattered particles back to the ISM rest system. Their LF there,
the maximum possible one, is $2\,\gamma_0^2$. A relativistic racket is an incredibly
efficient accelerator!

The $\Theta$ function in Eq.~(\ref{eq:source}), converted to a limit on energy,
(in $c=1$ units) is:
\begin{equation}
E_{\rm max} = 2\,\gamma_0^2\,M .
\label{eq:Emax}
\end{equation}
This is the key prediction: there must be a spectral feature in the different species of 
CRs at an energy of $\sim (2\,{\rm to}\,4)\!\times\! 10^6$ their mass. Since the distribution of
values of $\gamma_0$ is not quite a $\delta$ function and  
``elastic" scattering\footnote{``Plastic" may be more precise than ``elastic". 
The CR re-emission occurs \cite{DD2008}, in the CB rest frame, at a $\gamma_{\rm_{ CB}}\leq \gamma_0$, not 
affecting the upper limit of Eq.~(\ref{eq:Emax}).} is 
dominant only up to $E_{\rm max}$, the spectral feature is a knee.

 Based on Fermi's hypothesis that CRs are accelerated by moving shocks
of magnetized material, the conventional wisdom is that features in the CR 
spectra of individual nuclides ought to have
energies scaling with charge. But, at least for the knees' energies there is, to my knowledge,
no equivalent to the $2\,\gamma_0^2$ proportionality factor of Eq.~(\ref{eq:Emax}).

The expression $dF/d\gamma_{_{\rm CR}}\propto \gamma_{_{\rm CR}}^{-\beta_s}$
of Eq.~(\ref{eq:source}) is not yet a prediction for a CR spectrum. At energies
well below the ankle, CRs are confined and accumulated
by the Galaxy's magnetic field for a time, $\tau_{\rm conf}$,
 that depends
on their charge, Z, and momentum, $p$. An observed spectrum $F_{_{\rm CR}}$ and the
source spectrum $F$ are therefore related by:
\begin{eqnarray}
\!\!F_{_{\rm CR}}\propto \tau_{\rm conf}\,F,\;\;\;\tau_{\rm conf}=\tau_0\,
({\rm Z}\,p_0/p)^{\beta_{\rm conf}};\nonumber\\
\tau_0\sim (2-3)\times 10^7\,{\rm years},\;\;\;{\beta_{\rm conf}}\sim 0.6\pm 0.1,
\label{eq:conf}
\end{eqnarray}
\noindent
with $p_0\sim 1$ GeV and $\tau_0$ and ${\beta_{\rm conf}}$ estimated from
observations of astrophysical and solar plasmas and corroborated
by measurements of the relative abundances of secondary CR isotopes  \cite{CONFI}.

Recent measurements of the B/C ratio \cite{AMS} imply, at low rigidity ($R$) smaller 
values of $\beta_{\rm conf}$ than given in Eq.~(\ref{eq:conf}), see Fig.~2 in \cite{AMS}.
As theoretically expected, $\beta_{\rm conf}(R)$ flattens as $R$ increases.
The measured value at the highest-rigidity point ($R=860$ GV) is $.52\pm.13$. 
The rigidities of the knees discussed here are $\sim 10^3$ times larger,
and the primary CR spectra are adequately described by single power laws for more
than three orders of magnitude below their knees, see Fig.~(\ref{fig:OldKnees}).
It is therefore reasonable, as we do in what follows, to adopt $\beta_{\rm conf}=0.6$, 
compatible with the results of \cite{CONFI} and \cite{AMS}.

An inspection of Fig.(\ref{fig:NewKnees}) below indicates that a value of $\beta_{\rm conf}$ smaller than
0.6 would result in a slightly better description of the data. A reason to use an ``old" value of $\beta_{\rm conf}$ 
is that this paper deals with the predictions of an ``old" theory, rather than a description of recent data.

\subsection{CR abundances}

Let us pause to check whether we are on the right track. 
It is customary to discuss the composition of CR nuclei at a fixed
energy $E_{_{\rm A}}=1$ TeV.
This energy is relativistic ($p\simeq E$), below the 
corresponding knees for all A, and in the domain wherein the 
fluxes are well approximated by a power law with the 
index $\beta_{\rm th}\!=\!\beta_{s}\!+\!\beta_{\rm conf}\!\simeq\!2.77$, predicted by
combining Eqs.~(\ref{eq:source}) and (\ref{eq:conf}). 
Expressed in terms of energy ($E_{_{\rm A}}\!\propto\! {\rm A}\,\gamma$), 
read below the knee of Eq.~(\ref{eq:Emax})
and modified 
by confinement as in Eq.~(\ref{eq:conf}), Eq.~(\ref{eq:source}) becomes:
\begin{equation}
{dF_{\rm obs}\over dE_{_{\rm A}}}\propto {n}_{_{\rm A}}\,{\rm A}^{\beta_{\rm th}-1}
\,E_{_{\rm A}}^{-\beta_{\rm th}},\;\;\;
X_{_{\rm CR}}({\rm A})={{n}_{_{\rm A}}\over {n}_p}\,{\rm A}^{1.77},
\label{eq:compo}
\end{equation}
with ${n}_{_{\rm A}}$ an average ISM abundance and $X_{_{\rm CR}}({\rm A})$
the CR abundances relative to H, at fixed $E_{_{\rm A}}$  \cite{DD2008}.

\begin{figure}[]
\centering
\epsfig{file=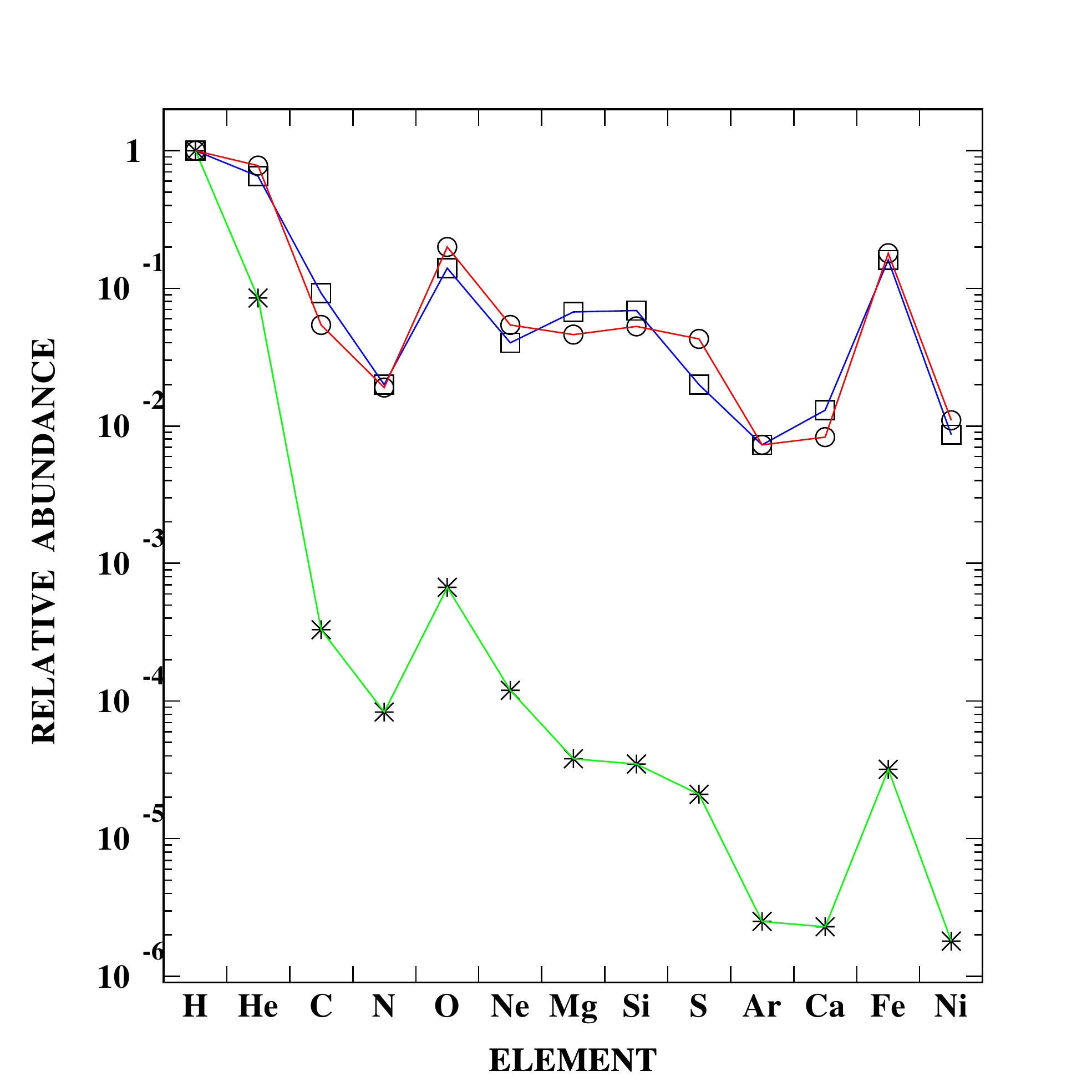,width=8cm}
\caption{The relative abundances of primary CR nuclei, from
H to Ni around 1 TeV \cite{BWS}. The stars (joined by green lines)
are solar-ISM abundances \cite{GS}. The circles (joined in red)
are the predictions, with input superbubble abundances. The
squares (joined in black) are the CR observations.}
\label{fig:abundances}
\end{figure}

The results of Eq.~(\ref{eq:compo}), for input ${n}_{_{\rm A}}$'s in the `superbubbles' wherein most
SNe occur, are shown in Fig.~(\ref{fig:abundances}). In these regions, the
abundances are a factor
$\sim\!3$ more `metallic' than solar.
The data
snugly reproduce the large enhancements of the heavy-CR relative abundances,
in comparison with solar or superbubble abundances
(e.g.~${\rm A}^{1.77}\!=\! 1242$ for Fe). 

Within the large uncertainties
of its priors (supernova rates, confinement time and volume) the
CB model accounts for the normalization of the all-particle CR flux, dominated by protons. 
With smaller uncertainties in the input priors, 
we have seen that the relative abundances of the various primary CR nuclei are fairly well reproduced.
Consequently, in comparing theory and observations of the 
CR fluxes of each given nuclide, we shall take the liberty of
fitting ``by eye" the overall normalization of the theory  to the data.

The CB model of CRs is a model of the origin and spectra of {\it primary} CRs.
Once one has an (alleged) understanding of the injection spectrum of primaries arriving not so far from the Earth, the understanding of secondaries is common to all models and it is relatively well understood, at least at the qualitative level and at all but the smallest rigidities. Secondaries are not a clear-cut tool to distinguish different models of the origin of CRs. 

The secondary to primary ratios (see, for instance, Fig. 1 of
\cite{AMS} for the B/C ratio) are seen --and theoretically understood \cite{JSM}-- 
to decrease very fast with energy.
At the energies of interest to the CR knees (three orders of magnitude above the highest measured rigidities, of order $2\times 10^3$ GV) the depletion of the primary CRs cannot be an effect greater than 
few tens of a percent. This is certainly negligible relative to the dominant uncertainties in the 
CB-model priors, such as the all-particle flux normalization.

\bigskip

\section{Back to the spectral knees}

A prior in discussing the position and shape of the knees is $D(\gamma_0)$, the distribution
of initial LFs of CBs, of which an example was given in Fig.~(\ref{fig:gammaDist}). Convoluted
with $D(\gamma_0)$ and as a function of energy, Eq.~(\ref{eq:source}) becomes:
\begin{eqnarray}
F_{_{\rm A}}=E_{_{\rm A}}^{-\beta_{\rm th}}\;{\rm Knee}(E_{_{\rm A}}),
\nonumber\\
\beta_{\rm th}=\beta_{s}+\beta_{\rm conf}\sim 2.77,
\nonumber\\
{\rm Knee(E_{_{\rm A}})}\equiv\int_{E_{_{\rm A}}/[2 M({\rm A})]}^\infty D(\gamma_0^2)\, d\gamma_0^2
\label{eq:fluxes}
\end{eqnarray}

\begin{figure}[]
\epsfig{file=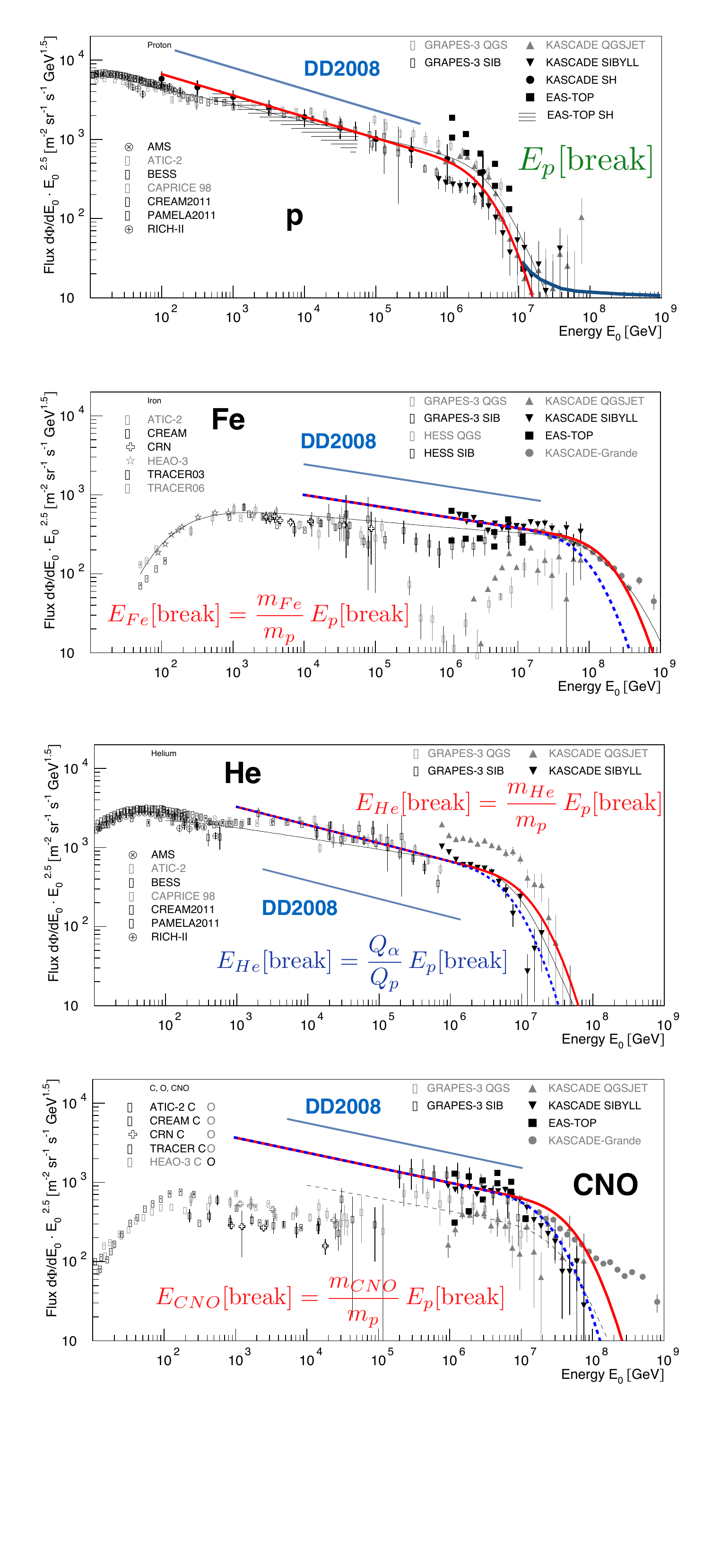,width=9cm}
\vspace{-2.5cm}
\caption{The spectra (times $E^{2.5}$) of primary CR nuclides: p, Fe, He and CNO \cite{Hor2}. 
The relative positions of their knees
are predicted by the CB model (red lines). Their exact absolute positions and shapes are obtained by adjusting a prior (the distribution $D(\gamma_0)$ of initial CB Lorentz factors) within its uncertainties. 
The dashed blue lines correspond to the hypothesis that the knees scale with charge, not mass.
The ``DD2008" lines are the expected spectral slopes \cite{DD2008}, to be continued to higher energies
in the absence of knees. The (red) CB-model results
 are totally satisfactory, but for the CNO group, whose spectrum as measured by KASKADE-Grande
 \cite{Hor2} is somewhat peculiar.}
\label{fig:NewKnees}
\end{figure}

The most recent data on the CR spectra of individual elements are shown in Fig.~(\ref{fig:NewKnees}).
The  CB-model curves are colored and correspond to a log-normal distribution: 
\begin{eqnarray}
D(x)={\rm Exp}(-[(x-x_0)/c]^2);\nonumber\\
x\equiv {\rm Log}_{10}[\gamma_0^2],\;\;x_0=6.3,\;\;c=0.5,
\label{eq:Dgamma}
\end{eqnarray}
\noindent
which results in the red curve that
satisfactorily describes the proton's knee (this function peaks  at a value 
of ${\rm Log}_{10}[\gamma_0]$ some 4\% larger than the prior
shown in Fig.~(\ref{fig:gammaDist}) and is about twice as large). 
The blue curve beyond the knee corresponds to a contribution, not needed for the current discussion,
of protons accelerated within the CBs \cite{DD2008} (for all CR nuclei, this contribution 
improves the agreement between theory and
data, as in Fig.~(\ref{fig:OldKnees}), at energies well below the knee).
The curves labeled DD2008 in Fig.~(\ref{fig:NewKnees}) 
correspond to the prediction of Eqs.(\ref{eq:fluxes})
with  $\rm Knee(E_{_{\rm A}})=1$, i.e.~no knee.

The next CR nuclide shown in Fig.~(\ref{fig:NewKnees}) is Fe since, should one trust
the recent KASKADE-Grande 
data \cite{Hor2} more than previous ones, they provide the best-measured knee.
The shape of the red curve is this time completely predicted, since the $D(\gamma_0)$ prior has
been chosen to describe the proton's knee.  Once again, the result is based on the simple kinematical
fact that in the CB model the knee positions scale with mass, not charge. The dotted blue curve
corresponds to the later case. Its exclusion is not as clear-cut as the figure seems to imply. One could have
fit $D(\gamma_0)$ to the Fe knee to predict the proton data. Since these do not appear to
be so precise, the exclusion of the generally assumed dependence on charge would have been 
a wee bit less convincing.



The data on He shown in Fig.~(\ref{fig:NewKnees}) display a much clearer knee than the older
ones in Fig.~(\ref{fig:OldKnees}). Once again, they do not establish a distinction between knees
scaling with charge or mass. Finally, the data on the CNO group also have a knee, whose shape makes
one wonder.

\section{The electron knee}

The mass ratio of Fe to protons is $\sim 56$, while their charge ratio is 26.
The relatively small ratio of these numbers ($\sim 2.15$) --combined with the errors in the data
and their spread-- are such that we could not establish a clear preference between charge and
mass in the positions of their respective spectral knees. A comparison between electrons and protons,
with a charge ratio of $[-]1$ and a mass ratio of $\sim 1832$, could prove decisive.

The spectral index for electrons of energies close to their putative knee is not 
$\beta_{\rm th}\sim 2.77$ as in Eq.~(\ref{eq:fluxes}) but $\beta_{\rm e}=\beta_{\rm th}+1$.
The reason \cite{DD2008} is that in this energy domain electrons (and positrons) efficiently lose energy by 
synchrotron radiation in the Galaxy's magnetic field and inverse Compton scattering on ambient 
photons\footnote{ This efficient energy loss does not imply that the electrons lose all of their 
energy in arriving to our planet from a typical supernova site. In the CB model CBs generate CRs 
along their trajectories, that typically extend well beyond the galactic disk.
We have not modeled in detail this very complex issue.}.
The knee function in Eq.~(\ref{eq:fluxes}) only requires the substitution of $M({\rm A})$ for $m_{\rm e}$.

The currently available relevant data are for the sum of $e^+$ and $e^-$ fluxes. 
The most conservative assumption is that positrons are CR secondaries, generated in
CR collisions with the ISM, producing pions (or kaons) with a decay chain 
$\pi^+\,({\rm or}\,K^+)\, \to \mu^+\, \nu$, $\mu^+ \to e^+\,\nu\,\bar\nu$. 
The same collisions generate secondary electrons with a similar energy distribution,
but in smaller quantities, due to the CR nuclei and their ISM targets being positively charged.
At a fixed energy the ratio of the secondary electron flux to the one of positrons ought
to be $\sim 0.74$, the measured $\mu^+/\mu^-$ ratio \cite{DaDa}.

In testing the prediction for the electron knee, I add to its spectrum the relatively small
contribution of secondary $e^+$ and $e^-$ fluxes, to be able to compare with the current data
up to the highest measured energies.
For the secondary $e^+$ flux I use the expressions in  \cite{DaDa}, which are conservative
(in the sense of the previous paragraph) and snuggly fit the data.

\begin{figure}[htbp]
\hspace{1.cm}
\includegraphics[width=.42\textwidth]{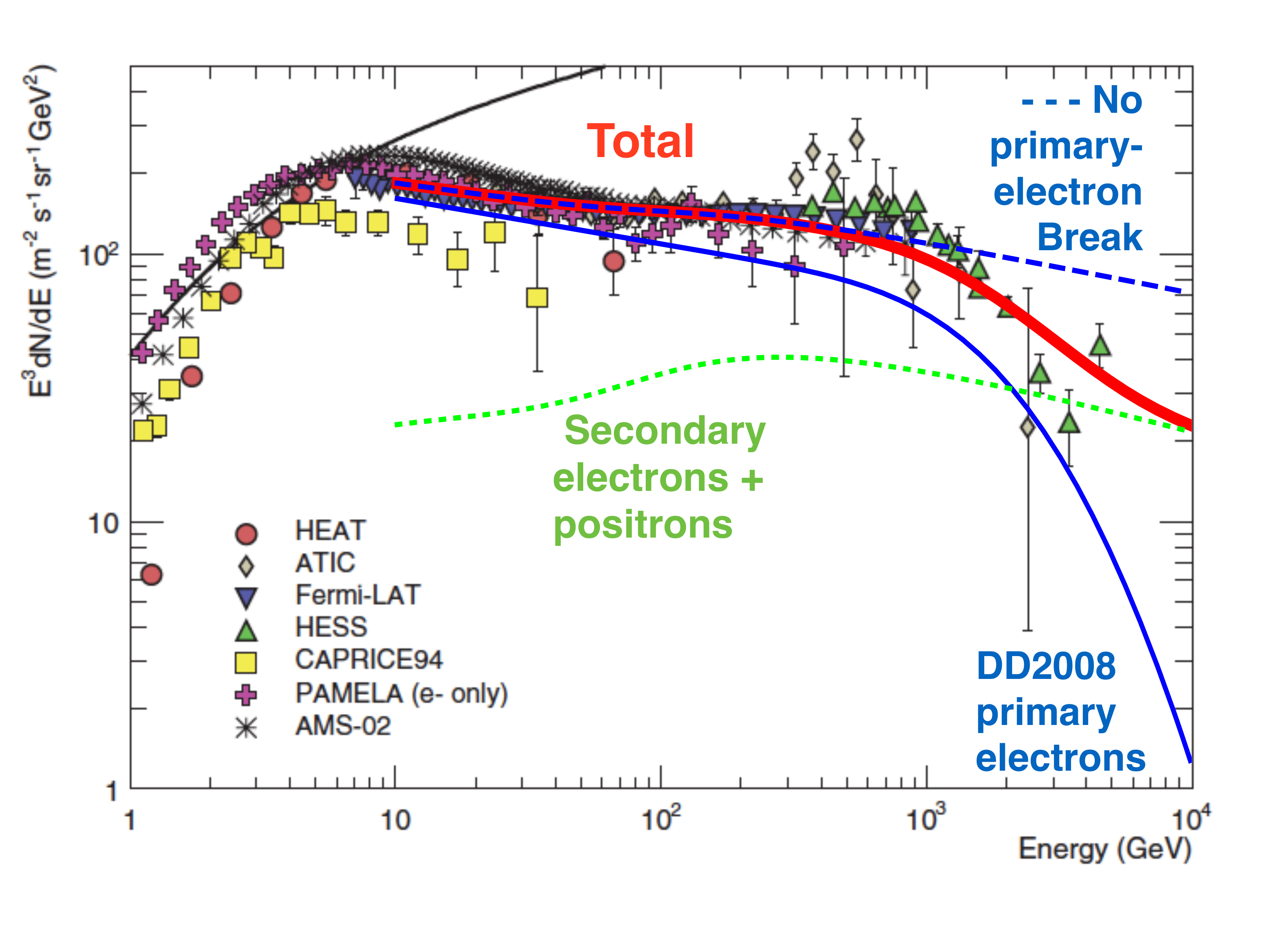}\\
\vspace{-0.8cm}
\hspace{-1.cm}
\includegraphics[width=.54\textwidth]{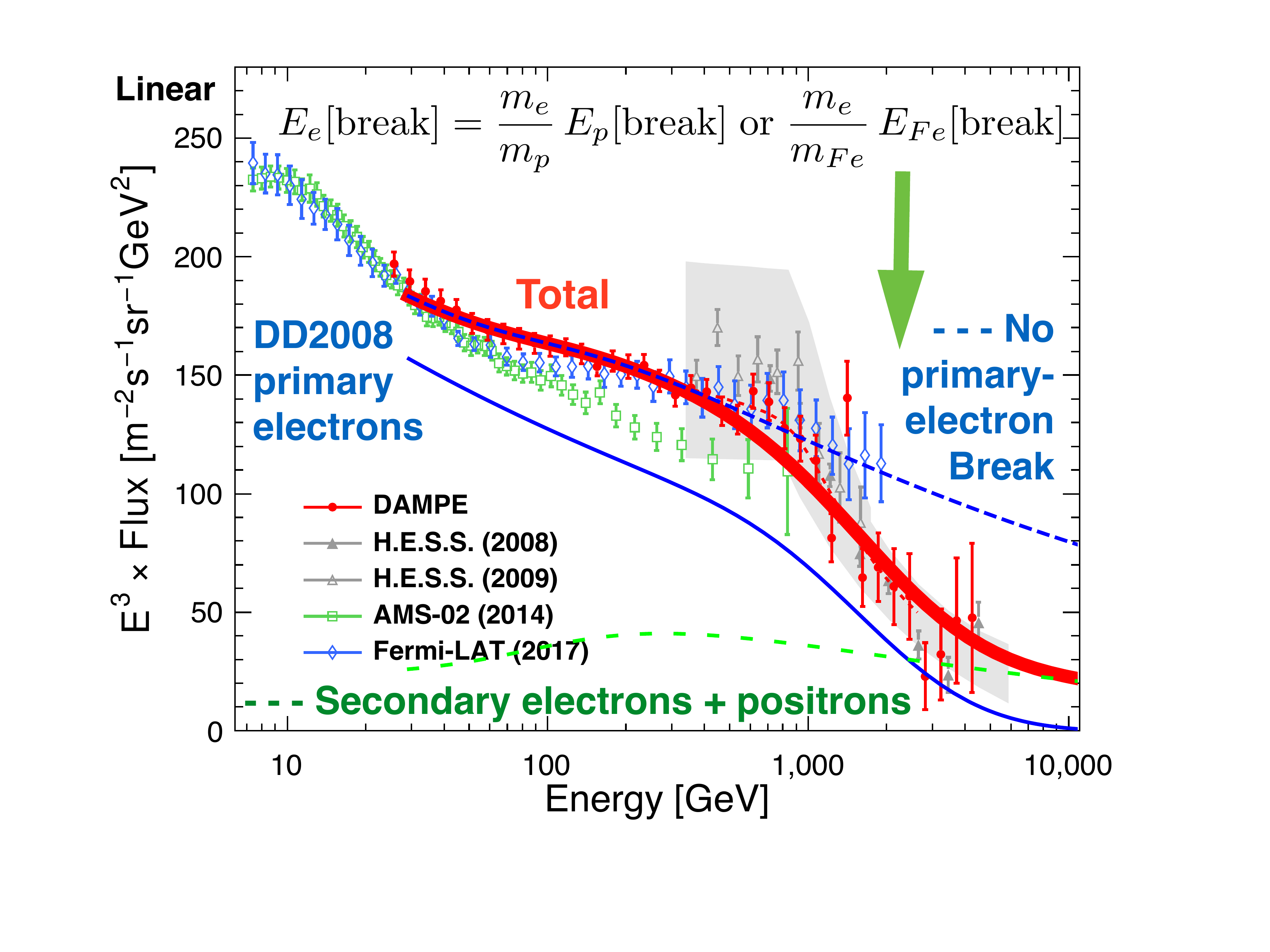}\\
\vspace{-1.cm}
\caption{Top: Relatively recent data on the $e^++\,e^-$ CR spectra (times $E^3$), on a 
log-log scale \cite{rpp}. The spectral knee predicted
by the CB model in the primary $e^-$ spectrum is clearly seen. Bottom: More recent data on a
linear-log scale \cite{Dampe}.
The knee is very conspicuosly observed in the DAMPE and HESS results, which reach the highest
energies. The dashed green curves are the secondary $e^++\,e^-$ CR spectra. The blue
curves are the spectra of primary electrons (with a fitted normalization). The dotted blue curves describe a
would-be theory with no knee. The thick red curve is the complete CB-model result.
  \label{fig:electrons}}
\end{figure}

Two sets of data and the CB-model fit are shown in the busy Fig.~(\ref{fig:electrons}). This is a 
(one-parameter) fit in the sense that we have not attempted to predict in the CB model the 
absolute normalization of the CR electron spectrum. The top of Fig.~(\ref{fig:electrons})
contains a relatively new set of data in a log-log plot, like previous figures.
The lower part of the figure is a linear-log plot, thus the (apparently)
dissimilar shapes of the differently colored
lines, which are precisely the same functions in the two plots.

The upper part of Fig.~(\ref{fig:electrons}) shows that there is indeed an electron knee at
the energy predicted by the CB model. Its shape is also compatible with the theoretical
prediction. The lower part of Fig.~(\ref{fig:electrons}) thickens the plot. The 
AMS-02 data are incompatible with the others and, since they do not reach the highest-measured energies,
cannot significantly distinguish a knee from its absence.  Finally, the data reaching
the highest energies, from HESS and DAMPE, agree with the presence of the primary electron
knee, with its predicted position and shape.
 


\section{Summary and Conclusions}

In the CB model there must be a knee in the spectra of primary CR nuclides and electrons at an energy of
(2 to 4) $\times \, 10^6$ times the particle's rest energy. A decade ago the prediction was only (succesfully)
testable for protons and --to some extent-- for Fe, whose spectral knee was already indirectly observable
as the ``second knee" in the all-particle spectrum \cite{DD2008}.

Recent data allow one to test the cited prediction. It turns out to be right for protons, He and Fe. 
The measurement errors, however, are insufficient to decide whether the observed knees scale with
mass or charge (the second choice would relate the relative positions of the knees, but does not predict
their absolute energies). 

Clearly, a decisive test would involve a measurement of the primary electron spectrum, since the
charge ratio of protons to electrons is extremely different from their mass ratio: $\sim 1832$;
not to speak of the mass ratio of Fe to electrons: five orders of magnitude!
There is not yet a measurement of the electron spectrum up to its predicted knee. But measurements
of the separate $e^+$ and $e^-$ spectra exist, up to energies a bit below that of the predicted
primary-electron knee. 

In a theory not invoking non-standard physics, the $e^+$ CRs are secondary,
and accompanied by a predictable amount of secondary $e^-$'s.  The 
individual lepton spectra are very well described in such a theory \cite{DaDa}.
 With its help, and the CB-model prediction
for the primary $e^-$ spectrum, I have argued that the CB-model's knee is observed. This
involves a modest extrapolation of the cited theory of secondary spectra, but there is
no reason to expect a break in these spectra\footnote{The AMS fit to their data \cite{AMS2} has an extremely
abrupt exponential cutoff beyond the measured energies. There is no standard-physics reason
to expect it.}, generated by collisions between higher-energy
CR nuclei and the ISM, and convoluted with the corresponding very broad spectra of 
secondary pions (and kaons) and the chain of their decay products. 
Moreover, the contribution of the secondary $e^++\,e^-$
to the total flux is, up to the $e^-$ knee, quite negligible, see Fig.~(\ref{fig:electrons}).

All in all, the CB-model's prediction of a knee in the CR spectra of primary hadrons
and electrons turns out to be correct. The only caveats are related to peculiarities of the data,
see the CNO spectrum of Fig.~(\ref{fig:NewKnees}) and the disagreements between
experiments in the lower part of Fig.~(\ref{fig:electrons}). Perhaps the correct conclusion
at this point would be the one attributed to Eddington: {\it Never trust an experiment until it has been 
confirmed by theory.}

\section{Added Note}
\begin{figure}[]
\epsfig{file=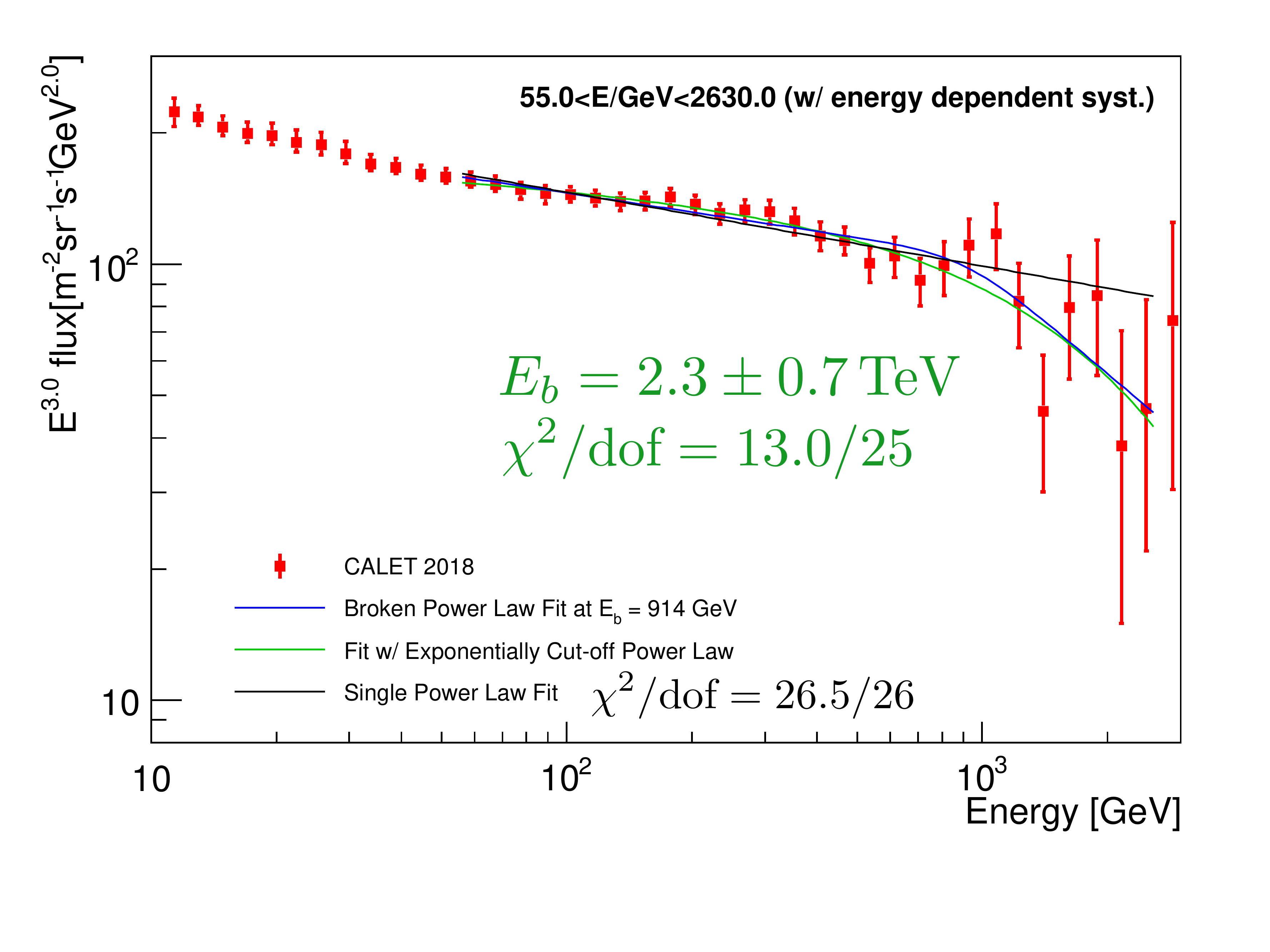,width=9cm}
\vspace{-1.5cm}
\caption{CALET data on the $e^+ +\,e^-$ spectrum. The black curve is a single power-law. The green curve
includes an exponential break.}
\label{fig:CALET}
\end{figure}

Right after the original version of this paper was posted,
relevant new data on the $e^+ +\,e^-$ spectrum were published by CALET \cite{CALET}.
They are shown in Fig.(\ref{fig:CALET}). Interestingly, they agree with the AMS data shown in the lower
panel of Fig.(\ref{fig:electrons}). But they extend to higher, knee-sensitive, energies.

The black single power-law fit of Fig.(\ref{fig:CALET}) has $\chi^2/\rm dof=26.5/26$. The green curve is
a fit with $\chi^2/\rm dof=13.0/25$ and an exponential break at $E_b=2.3\pm 0.7$ TeV. The only daringly
precise prediction for the position of the electron knee --also an exponential cutoff, as in Eqs.(\ref{eq:fluxes},
\ref{eq:Dgamma})-- is $E_b=2.3$ TeV \cite{ADRprecise}. This coincidence is intriguing, but the difference in
the fit qualities of the single-power-law and the exponentially-cutoff fits is, by itself, insufficient to justify
any strong claims.

\vspace{.5cm}
\noindent {\bf Acknowledgment:} A. De R\'ujula acknowledges that this project has received funding/support from the European UnionÕs Horizon 2020 research and innovation programme under the Marie Sklodowska-Curie grant agreement No 690575. I am particularly indebted to Shlomo Dado and Arnon Dar for discussions,
a long-time collaboration and for lending me their analytic expressions for the fluxes of CR positrons and
secondary electrons.


\vspace{.4 cm}
{\bf  Appendix: The CB model of GRBs and CRs}
\vspace{.1 cm}

\begin{figure}[]
\centering
\vspace{.5cm}
\epsfig{file=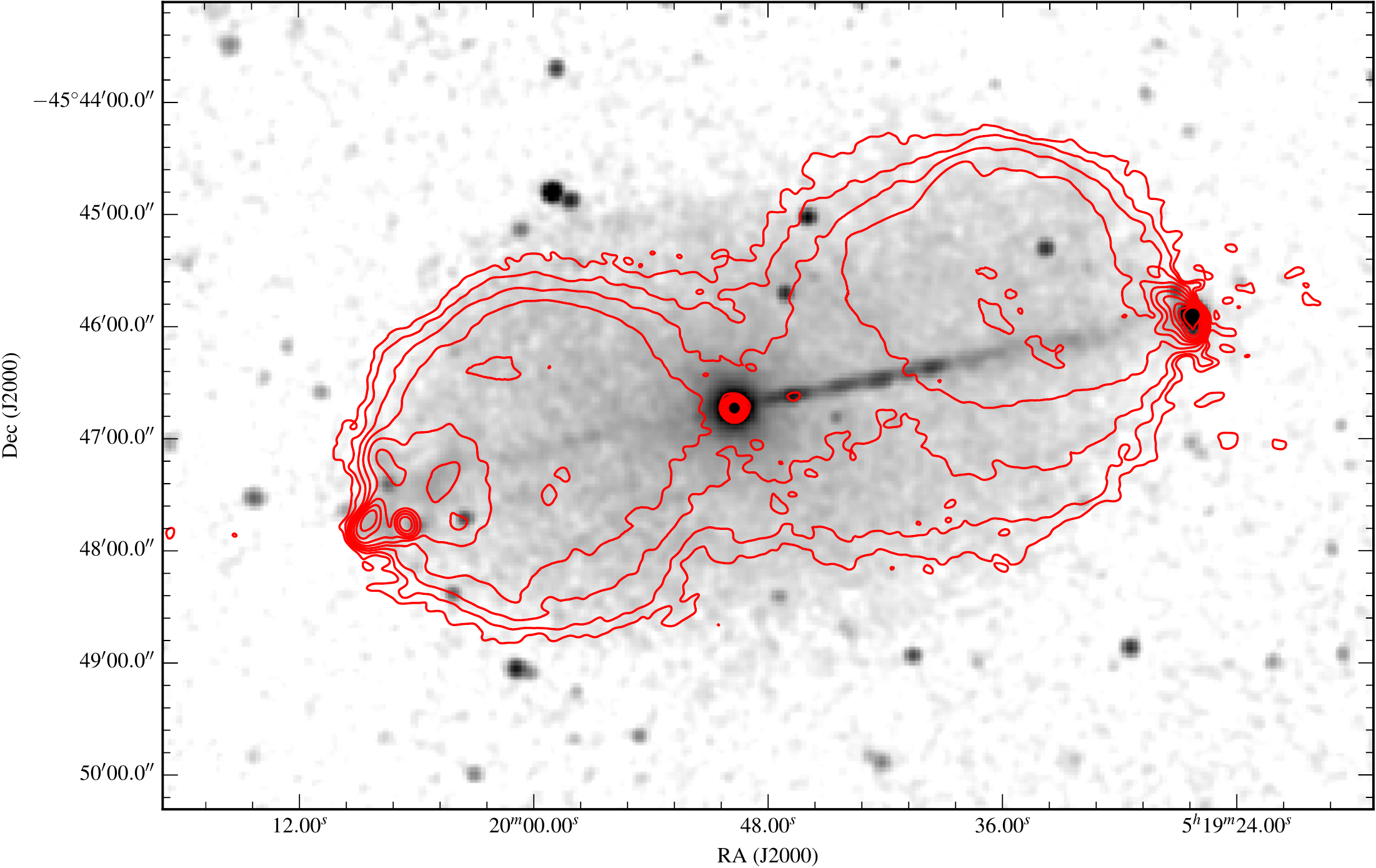,width=8.cm}
\caption{The quasar Pictor A. A superposition of an X-ray image and the (red)
contours of the radio emission
\cite{Pictor}.}
\label{fig:PictorA}
\end{figure}

Jets are emitted by many astrophysical systems, such as Pictor A,
shown in Fig.~(\ref{fig:PictorA}). Its active
galactic nucleus is discontinuously spitting something that, seen in X-rays, does not
appear to expand sideways before it stops and blows up, having by then
travelled almost $10^6$ light years. Many such 
systems have been observed. They are relativistic: the Lorentz factors (LFs)
$\gamma\!\equiv\! E/(mc^2)$ of their ejecta are of ${\cal{O}}(10)$.
The mechanism responsible for these ejections,
due to episodes of violent accretion into a very
massive black hole, is not understood in detail.

The radio signal in Fig.~(\ref{fig:PictorA}) is the synchrotron radiation of
Ôcosmic-rayÕ electrons \cite{Pictor}. Electrons and nuclei were scattered
by the CBs of Pictor A, which encountered them at rest in the
intergalactic medium, kicking them up to high energies.
Thereafter, these particles diffuse in the ambient
magnetic fields (that they contribute to generate) and the electrons radiate.

In our galaxy there are `micro-quasars', whose central black
hole's mass is a few $M_\odot$. 
The best studied \cite{GRS}
is GRS 1915+105.
A-periodically, about once a month, this object emits two
oppositely directed {\it cannonballs}, travelling at $v\sim 0.92\, c$.
When this happens, the continuous X-ray emissions
 ---attributed to an unstable accretion disk--- temporarily decrease.
Atomic lines of many elements have been seen
in the CBs of $\mu$-quasar SS 433 \cite{SS433}. Thus, at least in this case, the
ejecta are made of ordinary matter, and not of a fancier substance,
such as $e^+e^-$ pairs.

The `cannon' of the CB model is analogous to the ones
of quasars and $\mu$-quasars.
In the {core-collapse} responsible for a stripped-envelope
SNIc event, due to the parent star's
rotation, an accretion disk  is produced around
the newly-born compact object,  by stellar material originally
close to the surface of the imploding core, or by more distant stellar matter
falling back after the shock's passage. 
 A CB made of {\it ordinary-matter plasma} is emitted, as
in microquasars, when part of the accretion disk
falls abruptly onto the compact object. {\it Long-duration} GRBs  
and {\it non-solar} CRs are produced by these jetted CBs.

\begin{figure}
\hskip 2truecm
\begin{center}
{\epsfig{file=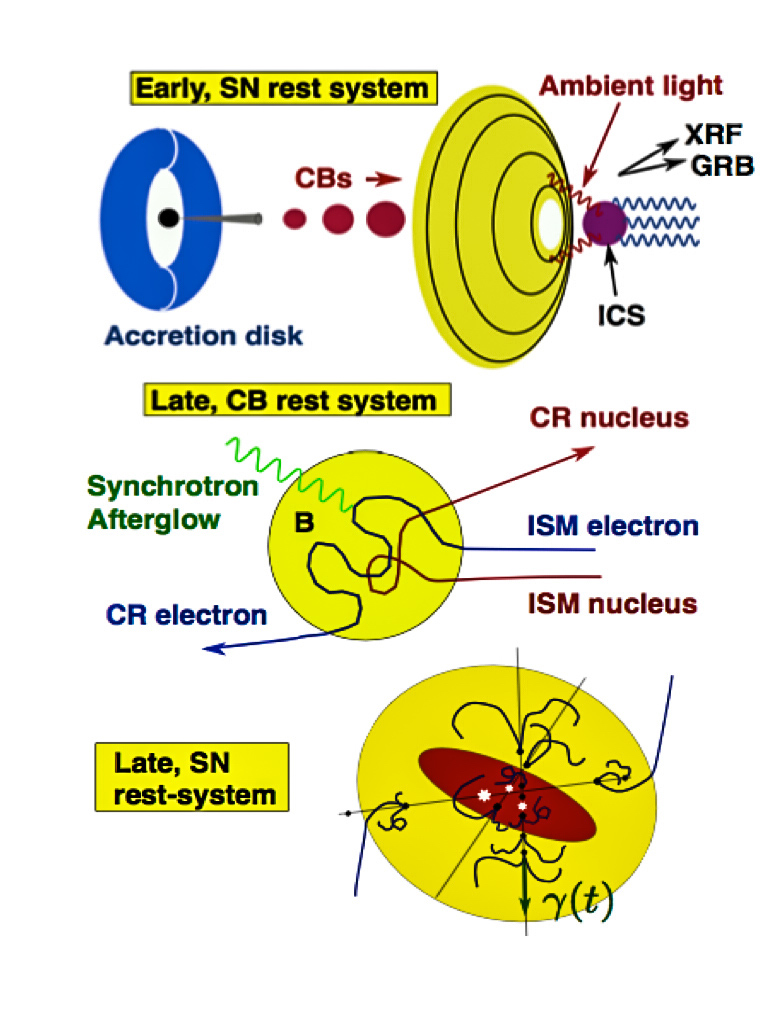, width=6.5cm}}
\end{center}
\vspace{-1.cm}
\caption{The CB model
of GRBs, XRFs and CRs. A core-collapse SN results in
a (black) compact object, a (blue) fast-rotating torus of non-ejected
material and a (yellow) shell of non-relativistic ejecta. 
Matter (not shown) episodically accreting
into the central object produces
two  collimated beams of (brown) CBs;  some of
the `Northern' ones are depicted. As the CBs move through
the {\it glory} of non-radial
 light surrounding the star, they forward Compton up-scatter
its photons to GRB or XRF energies, depending on how close
the line of sight is to the CBs' direction. 
Each CB produces a GRB or XRF `pulse'. Later, a CB gathers and
scatters ISM particles, which ``collide" with its inner magnetic field. 
When re-emitted, these particles are 
boosted by the CB's motion: they have become
CRs. The synchrotron radiation of the gathered electrons is the late AG
of the GRB or XRF. As a CB's
collisions with the ISM
slow it down, it generates CRs along its
trajectory, in the galaxy and its halo. 
These CRs diffuse thereafter in the
local magnetic fields.}
\label{fig:CB}
\end{figure}

A summary of the CB model of GRBs and  XRFs is given in Fig.~\ref{fig:CB}. 
The {\it `inverse' Compton scattering} (ICS) of light by electrons within a CB  
produces a highly forward-collimated beam of higher-energy photons.
The target light is in a temporary reservoir: the {\it glory}, 
an ÒechoÓ (or ambient)
light from the SN, permeating the Òwind-fedÓ circumburst density profile, previously ionized
by the early extreme UV flash accompanying a SN explosion, or by the enhanced UV emission
that precedes it.

Seen close to the CB's direction of motion, the beam of $\gamma$-rays is
a pulse of a GRB. Not so close, it is the pulse of an XRF. To agree with 
observations, CBs must be launched with LFs, $\gamma_0\!\sim\!10^3$,
and baryon numbers $N_{_{\rm B}}\!=\!{\cal{O}}(10^{50})$, corresponding to
$\sim\!1/2$ of the mass of Mercury, a miserable $\sim\!10^{-7}\,M_\odot$.

The simple kinematics describing a narrow beam of GRB or XRF photons --viewed at different
angles-- suffice to predict all observed correlations between pairs of prompt 
observables, e.g.~photon
fluence, energy fluence, peak intensity and luminosity, photon energy at peak 
intensity or luminosity, and pulse duration. The correlations are tightly obeyed,
indicating that GRBs are moderately standard candles (with ``absolute" properties
varying over a couple of orders of magnitude) while the observer's angle makes
their apparent properties vary over very many orders of magnitude  \cite{corr}. 
Double and triple correralions between GRB observables and the ``break time"
of afterglows are also in excellent agreement with the CB model \cite{corr1}.
Similarly
simple kinematics explain the positions of the knees in CR spectra.
The shapes of GRB pulses and their spectrum are also neatly explained by 
ICS of glory light \cite{DD}.

In its long journey through its host galaxy, a CB encounters
the constituens of the ISM, previously ionized by the GRB's $\gamma$-rays.
The merger of two plasmas (the ISM's and CB's constituency) at a large 
relative LF generates a CB's turbulent inner magnetic field, assumed to be in energy equipatition
with the kinetic energy of the entering ISM particles \cite{DD2008}. All this is corroborated by
simulations of plasma mergers \cite{merger}. CRs and the galactic
magnetic fields also have similar energy densities.
GRBs and XRFs have long-lasting {\it `afterglows'} (AGs). The CB model acounts
for them as synchrotron radiation from the ambient electrons swept in and
accelerated within the CBs, predicting the correct fluences, AG light curves and spectra
\cite{AGoptical, DDX}.

The only obstacle still separating the CB model from a complete theory of GRBs is
the theoretical understanding of the CBsÕ ejection mechanism in SN explosions.
Otherwise the CB model correctly describes all known properties of GRBs and XRFs.
But, perhaps more significantly, the model also resulted in remarkable predictions:

\vspace{.2cm}
{\bf The SN-GRB association}

GRB 980425 was {\it `associated'} with the supernova SN1998bw: within directional 
errors and within a timing uncertainty of $\sim\!1$ day, they coincided. The 
luminosity of a 1998bw-like SN peaks at $\sim 15\,(1+z)$ days. 
The SN light competes at that time and frequency with the AG of its 
GRB, and it is not always easily detectable.  {\it Iff} one has a predictive theory
of AGs, one may test whether GRBs are associated with  `standard torch' SNe, 
akin to SN1998bw, `transported' to the GRBs' redshifts. 
The test was already conclusive (to us) in 2001 \cite{AGoptical}.
One could even {\it foretell  the date} in which a GRB's SN 
would be discovered. For example, GRB 030329 was so
`very near' at $z\!=\!0.168$, that we could not resist posting such a daring
prediction \cite{SN030329} during the first few days of AG observations. 
The prediction turned out to be right.
The spectrum of this SN was very well measured and seen to coincide snugly with
that of SN1998bw. This is why the SN/GRB association ceased to be doubted.

\vspace{.2cm}
{\bf The AG light curves}

Swift has established a {\it canonical behaviour} of the X-ray and optical AGs of a 
large fraction of GRBs. The X-ray fluence decreases very fast from a
`prompt' maximum. It subsequently turns into a `plateau'. After a time of
${\cal{O}}(1$d), the fluence bends (has an achromatic `break', in the usual parlance)
and steepens to a power-decline. 
Although all this was considered a surprise, it was not \cite{canonical}. 
Even the good old GRB 980425, the first to be clearly
associated with a SN, sketched a canonical X-ray light curve, 
with what we called a `plateau' \cite{AGoptical}.
Dozens of X-ray and optical AGs have been shown to be correctly
described by the CB model \cite{AGoptical, DDX}.

\vspace{.2cm}
{\bf The superluminal motion}

Only in two SN explosions that took place close enough, the
CBs were in practice observable.  One case
was SN1987A, located in the LMC,
whose approaching and receding CBs were
photographed \cite{Costas}.
The other case was SN2003dh, associated with GRB030329,
at $z=0.1685$. In the CB model interpretation,
its two approaching CBs were first `seen', and fit,
as the two-peak $\gamma$-ray light curve 
and the two-shoulder AG.
This allowed us 
to estimate the time-varying angle of their apparent superluminal
motion in the sky \cite{SLum030329}. Two sources or `components'
were indeed clearly seen in radio observations
at a certain date, coincident
with an optical AG rebrightening. We claim
that the data agree with our expectations\footnote{The
size of a CB is small enough to expect its radio image to
scintillate, arguably more than observed \cite{Taylor}.
Admittedly, we only realized a posteriori that the ISM electrons a CB
scatters, synchrotron-radiating in the ambient magnetic field, would
significantly contribute at radio frequencies, somewhat blurring the 
CBs' radio image \cite{SLum030329}. Also, during the integration time 
of a radio observation the CBs would move in the sky, 
obliterating the scintillations \cite{NS}.},  including 
the predicted inter-CB separation \cite{SLum030329}.
The observers claimed the contrary, though the 
evidence for the weaker `second component' is $>20\sigma$.
They report \cite{Taylor} that this component is 
`not expected in the standard model'.

\vspace{.2cm}
 {\bf The GRB's $\gamma$-ray polarization}

Earliest but not least \cite{SD,DDDpol}. Let a CB launched with a LF $\gamma_0$ be seen at an angle $\theta$ 
from its jetted direction. The observed $\gamma$-rays, having been Compton up-scattered,
have a polarization $\Pi\!\approx\! 2\,\gamma_0^2\,\theta^2/(1+\gamma_0^4\,\theta^4)$.
This vanishes on axis, is nearly 100\% for the most probable viewing angle ($\theta\!\sim\! 1/\gamma_0$) and
$>\! 47$\% for $2/\gamma_0\!>\!\theta>1/(2\,\gamma_0)$. All measured GRB polarizations \cite{pol1}
are $>\!47$\%, but two, 930131 and 100826A, whose polarizations are
also incompatible with $\Pi=0$, the expectation for synchrotron radiation of electrons
in a non-structured magnetic field.

\end{document}